\documentclass[11pt,a4paper]{article}%
\usepackage{amsfonts}
\usepackage{amsmath}
\usepackage{amssymb}
\usepackage[margin=1in,a4paper]{geometry}
\usepackage[sc,center,compact,noindentafter,medium]{titlesec}
\usepackage[onehalfspacing,nodisplayskipstretch]{setspace}
\usepackage{graphicx}
\usepackage{color}%
\definecolor{darkblue}{rgb}{.2, 0.2,.8}
\definecolor{darkgreen}{rgb}{0,0.5,0.3}
\definecolor{darkred}{rgb}{.8, .1,.1}

\providecommand{\U}[1]{\protect\rule{.1in}{.1in}}

\newtheorem{theorem}{\normalfont\scshape Theorem}[section]
\newtheorem{corollary}{\normalfont\scshape Corollary}[section]
\newtheorem{lemma}{\normalfont\scshape Lemma}[section]

\expandafter
\def \expandafter \normalsize \expandafter{\normalsize \setlength \abovedisplayskip{10pt plus 2pt minus 7pt}}
\expandafter
\def \expandafter \normalsize \expandafter{\normalsize \setlength \abovedisplayshortskip{0pt plus 2pt}}
\expandafter
\def \expandafter \normalsize \expandafter{\normalsize \setlength \belowdisplayskip{10pt plus 2pt minus 7pt}}
\expandafter
\def \expandafter \normalsize \expandafter{\normalsize \setlength \belowdisplayshortskip{5pt plus 2pt minus 3pt}}

\numberwithin{equation}{section}
\graphicspath{{C:/temp/}}
\begin{document}

\title{ }

\begin{center}
{\LARGE \textsc{The Econometrics of Financial Duration Modeling}}%

\renewcommand{\thefootnote}{}
\footnote{
\hspace{-7.2mm}
$^{a}%
$Department of Economics, University of Bologna, Italy and Department of Economics, University of Exeter, UK.
\newline$^{b}%
$Department of Mathematical Sciences, University of Copenhagen, Denmark.
\newline$^{c}$Department of Economics, University of Copenhagen, Denmark.
\newline Correspondence to: Giuseppe Cavaliere, Department of
Economics, University of Bologna, email giuseppe.cavaliere@unibo.it.
}
\addtocounter{footnote}{-1}
\renewcommand{\thefootnote}{\arabic{footnote}}%
{\normalsize \vspace{0.1cm} }

{\large \textsc{Giuseppe Cavaliere}}$^{a}${\large \textsc{, Thomas Mikosch}%
}$^{b}$, {\large \textsc{Anders Rahbek}}$^{c}$

{\large \textsc{and Frederik Vilandt}}$^{c}${\normalsize \vspace{0.2cm}%
\vspace{0.2cm}}

November 21, 2022{\normalsize \vspace{0.2cm}\vspace{0.2cm}}

\textsc{Abstract\vspace{-0.15cm}}
\end{center}

We establish new results for estimation and inference in financial durations
models, where events are observed over a given time span, such as a trading
day, or a week. For the classical autoregressive conditional duration (ACD)
models by Engle and Russell (1998, \emph{Econometrica }66, 1127--1162), we
show that the large sample behavior of likelihood estimators is highly
sensitive to the tail behavior of the financial durations. In particular, even
under stationarity, asymptotic normality breaks down for tail indices smaller
than one or, equivalently, when the clustering behaviour of the observed
events is such that the unconditional distribution of the durations has no
finite mean. Instead, we find that estimators are mixed Gaussian and have
non-standard rates of convergence. The results are based on exploiting the
crucial fact that for duration data the number of observations within any
given time span is random. Our results apply to general econometric models
where the number of observed events is random. \bigskip


\bigskip

\section{Introduction}

\label{sec intro}In the seminal papers by Engle and Russell (1998) and Engle
(2000), autoregressive conditional duration (ACD) models were introduced for
modeling durations, or waiting times, between financial events, and to analyze
liquidity in financial markets. Financial events are observed over a given
period of time, such as a (trading) day, a week, or a year; hence, both the
size and the number of durations are random variables. As we
demonstrate\textbf{,} the randomness of the number of events has a major
impact on asymptotics and inference in dynamic duration models. Moreover, as
detailed below, existing results cover alone the case where the number of
events is non-random and therefore are not applicable to estimation of ACD
models over a given time span. In this paper, we provide the missing
asymptotic analysis for likelihood-based estimators. We specifically show that
the randomness of the number of events plays a crucial role and leads to a new
distributional theory (at non-standard rates of convergence)\ for likelihood
estimators and related test statistics. The derivation of these novel results
requires non-standard asymptotic arguments, combining new results on the tail
behaviour of the durations with renewal theory.

The ACD models are by now quite popular in financial econometrics; see e.g.
Hautsch (2012) and Fernandes, Medeiros and Veiga (2016) for applications and
theory in the context of high-frequency data and Pacurar (2008) for a general
survey. Applications of dynamic duration models such as the ACD\ are
extensively used also in different areas of economics; see e.g. Hamilton and
Jord\`{a} (2002) or Aquilina, Budish and O'Neill (2022).

Let $[0,T]$ denote the observation period, where we observe $n$ event times
$\left\{  t_{i}\right\}  _{i=1}^{n}$, $0<t_{1}<t_{2}<\cdots<t_{n}\leq T$, with
corresponding durations $x_{i}=t_{i}-t_{i-1}$, $i=1,...,n$, $t_{0}=0$. As
noted in Engle and Russell (1998), $n$ is the realization at time $t=T$ of the
stochastic counting process $N_{t}$, $t\geq0$, given by
\begin{equation}
N_{t}=\#\big\{k\geq1:t_{k}=x_{1}+\cdots+x_{k}\leq
t\big\}.\label{eq: random summation index}%
\end{equation}
In particular, the number of events $N_{T}$, in the observation period
$[0,T]$, is as mentioned a \emph{random variable}.

The most known dynamic duration model is the ACD\ of Engle and Russell (1998)
which in its simplest version (ACD\ of order one) is given by
\begin{equation}
x_{i}=\psi_{i}\left(  \theta\right)  \varepsilon_{i},\text{ \ \ }\psi
_{i}\left(  \theta\right)  =\omega+\alpha x_{i-1},\quad i=1,\ldots,N_{T}\,,
\label{eq: EACD model}%
\end{equation}
where $\theta=(\omega,\alpha)^{\prime}$ and $\psi_{i}(\theta)$ is the
conditional (duration) rate of the \emph{i}th waiting time $x_{i}$, i.e.,
conditional on $\mathcal{F}_{i-1}=\sigma(x_{i-1},x_{i-2},\ldots)$. The
innovations $\{\varepsilon_{i}\}$ are assumed i.i.d., strictly positive, with
unit mean, $\mathbb{E}[\varepsilon_{i}]=1$. If $\varepsilon_{i}$ is
exponentially distributed this is referred to as \emph{exponential} ACD (EACD).

With parameters $\theta=(\omega,\alpha)^{\prime}$, for $\omega>0$, $\alpha
\geq0$, and observation period $[0,T]$, the EACD log-likelihood function is
{given by}
\begin{equation}
L_{T}\left(  \theta\right)  =-\sum_{i=1}^{N_{T}}\left[  \log\psi_{i}\left(
\theta\right)  +\frac{x_{i}}{\psi_{i}\left(  \theta\right)  }\right]
\,,\qquad T\geq0\,. \label{eq: EACD likelihood}%
\end{equation}
Then $\hat{\theta}_{T}=\arg\max_{\theta}L_{T}\left(  \theta\right)  $ denotes
the maximum likelihood estimator (MLE) of $\theta$ in the case of i.i.d.
exponentially distributed $\{\varepsilon_{i}\}$, otherwise we refer to it as a
quasi maximum likelihood estimator (QMLE).

Engle and Russell (1998) note that the log-likelihood function in
(\ref{eq: EACD likelihood}) has the same form as the log-likelihood function
for the autoregressive conditional heteroskedastic (ARCH) model with Gaussian
innovations, and quote standard asymptotic theory from ARCH models in Lee and
Hansen (1996); see also Fernandes and Grummig (2006), Hautsch\ (2012, Theorem
5.2), Allen, Felix, McAleer and Peiris (2008) and Sin (2014) for a similar
approach to inference. Importantly, this approach treats $N_{T}$ as
\emph{deterministic}; that is, sampling is by number of durations and not over
a fixed, predetermined observation period $[0,T]$. Importantly, the results
for deterministic $N_{T}$ cannot be applied to the case of random $N_{T}$, as
analyzed here.

To give an idea of the difference in arguments between the two different
sampling schemes, a key insight is that the fact that the number of
observations $N_{T}$ is random implies that classical laws of large numbers
(LLNs) and central limit theorems (CLTs) are no longer directly applicable to
likelihood-related quantities such as score and information. For instance, it
is known from renewal process theory (see e.g. Gut, 2009) that $N_{T}$
$\rightarrow\infty$ is not sufficient for the LLN\ or the CLT to apply to
series of the form $Y_{T}=\sum_{i=1}^{N_{T}}\xi_{i}$, where both $N_{T}$ and
the random variables $\{\xi_{i}\}$ are defined in terms of the durations
$\{x_{i}\}$; such series appear repeatedly in the asymptotic theory for ACD.
In contrast to the deterministic $N_{T}$ case, the large sample behaviour of
$Y_{T}$ is intimately related to the large sample properties of the counts
$N_{T}$, which, again, depends on the tail properties (and existence of
moments) of the marginal distribution of the stationary and ergodic duration
$x_{i}$. Such dependence leads to a novel asymptotic theory, based on
non-standard arguments.

Specifically, as we show in this paper, the asymptotic theory for the
MLE\ crucially depends on the tail behavior and existence of moments for the
ergodic and stationary durations $\{x_{i}\}$, with the tail behaviour
characterized by the tail index $\kappa>0$ of the marginal distribution of
$x_{i}$; $P(x_{i}>z)\sim c_{\kappa}z^{-\kappa}$ as $z\rightarrow\infty$ for
some constant $c_{\kappa}>0$. We show that while asymptotic normality holds
for $\kappa>1$,\thinspace or equivalently, when the durations have finite
mean, asymptotic normality breaks down for $\kappa<1$. This is a crucial fact,
given that a wide range of tail indices is witnessed in empirical applications
on duration data. Thus, for example, Hill estimation of $\kappa$ yields
$\hat{\kappa}=2.1>2$ for the IBM\ transaction data analyzed in Engle and
Russell (1998) and $\hat{\kappa}=2.5>2$ on the durations between tweets in
Cavaliere, Lu, Rahbek and St\ae rk-\O stergaard (2022). Moreover, $\hat
{\kappa}=1.4\in\left(  1,2\right)  $ for the DJIA\ data from Embrechts,
Liniger and Lin (2011), while $\hat{\kappa}=0.7<1$ on SPY\ transaction data
over a single trading day (2019:07:31). Notably, while asymptotic normality
holds for $\kappa>1$, the Gaussian finite sample approximation is poor for the
case of infinite variance $\kappa<2$ and indeed invalid for the case of
infinite mean where $\kappa<1$.

A preview of our results is as follows. In classic settings, with $\{x_{i}\}$
i.i.d. with finite mean, the number of events per unit of time $N_{T}/T$
converges in probability to a strictly positive constant, in which case LLNs
and CLTs for $\sum_{i=1}^{N_{T}}\xi_{i}$ can usually be verified;\textbf{ }see
e.g. Gut (2009) for a survey. In the ACD\ setting, whether this holds depends
on the tail index $\kappa$. On the one hand, if $\kappa>1$, hence
$\mathbb{E}[x_{i}]=\mu\in(0,\infty)$, and $N_{T}/T$ is such that
$N_{T}/T=1/\mu+o(1)$ a.s. However, even in this simpler case, existing
(renewal) theory does not include stationary and ergodic $x_{i}$, and we
provide the needed extensions to the theory here. On the other hand, if
$\kappa<1$, hence $\mathbb{E}[x_{i}]=\infty$, then $N_{T}/T$ converges
(a.s.)\ to zero as $T\rightarrow\infty$ and neither the classic LLN nor the
CLT apply to $\sum_{i=1}^{N_{T}}\xi_{i}$. New tools are required for the
asymptotic theory and, in particular,\textbf{ }we establish the novel result
that $N_{T}/T^{\kappa}$ converges in distribution to a random variable with an
unfamiliar distribution, and for which we provide an explicit expression in
terms of a $\kappa$-stable random variable.

These convergence results for $N_{T}$ are essential for establishing the
asymptotic distribution of the QMLE. Specifically, we show that, provided
$\mu=\mathbb{E}[x_{i}]<\infty$, $\hat{\theta}_{T}-\theta_{0}$ (with
$\theta_{0}$ denoting the true value) is indeed asymptotically Gaussian when
normalized by the standard deterministic $\sqrt{T}$-rate. However, while
$\mathbb{E}[x_{i}]<\infty$ is indeed sufficient for $\sqrt{T}$-convergence to
the Gaussian distribution, the quality of the Gaussian approximation in finite
samples is demonstrated to be very poor when $\mathbb{E}[x_{i}^{2}]=\infty$,
or $\kappa<2$, and deteriorating as the tail index $\kappa$ approaches one.
Hence even when $\mathbb{E}[x_{i}]<\infty$ these results question the
usefulness of the $\sqrt{T}$-Gaussian approximation for likelihood estimators
in ACD models. In the case $\kappa<1$, the fact that $N_{T}/T^{\kappa}$
converges in distribution -- and not in probability -- to a non-standard
random variable implies that the derivation of the limiting distribution of
$\hat{\theta}_{T}-\theta_{0}$ is non-standard. In particular we show that the
information is random in the limit, and that this results in a limiting mixed
Gaussian distribution of $\hat{\theta}_{T}-\theta_{0}$ with a convergence rate
which depends on the value of tail index $\kappa<1$. A further, novel result
that follows from our results is that the $t$ ratio for
(univariate)\ hypotheses on $\theta$ is asymptotically normal provided
$\kappa>1$ or $\kappa<1$. The local power function of the test, however,
crucially depends on $\kappa$. The case $\kappa=1$ is not covered by our
theorem, and hence particular attention should be paid to applications where
estimated parameters are close to the boundary case $Ex_{i}=\infty$.

To sum up, our results show that, in contrast to ARCH\ models where the
marginal distribution of the data does not play any role in the asymptotic
theory, for ACD\ models this is indeed crucial, as the tail index of the
duration determines the speed of convergence of the estimators as well as
their asymptotic distribution. As already mentioned this is of empirical
relevance, as both the case of infinite and finite mean durations ($\kappa<1$
and $\kappa>1$, respectively) are found in applications. Moreover, our
findings are not specific to ACD models, but apply to general econometric
method where the number of observations over a given time span needs being
treated as a random process; see also Section \ref{sec discussion}.

The paper is structured as follows. In Section~\ref{sec: prelim} we discuss
the tail behavior of ACD processes, and provide new results for the related
counting process $N_{T},$ $T\geq0$. In Section~\ref{sec asymptotics} we
present the main asymptotic theory. A discussion about the implications for
inference and some concluding remarks are given in Section
\ref{sec discussion}. All proofs are provided in the Appendix. In the
following, `$\overset{p}{\rightarrow}$', `$\overset{\mathrm{a.s.}}%
{\rightarrow}$', and `$\overset{d}{\rightarrow}$' refer to convergence in
probability, almost surely and in distribution, respectively, in all cases
when $T\rightarrow\infty$. A generic element of a strictly stationary sequence
$\{y_{i}\}$ is denoted by $y$.

\section{Preliminaries\label{sec: prelim}}

In this section we derive the required results on the tail properties of the
durations and on the asymptotic behaviour of the random number of durations
$N_{T}$. Results of this kind are neither present nor required in the
classical ARCH case, where $N_{T}$ is deterministic.

\subsection{Tail behavior of the ACD\label{sec tails of ACD}}

We consider the sequence\ $x_{i}=\psi_{i}\varepsilon_{i}$, $i\in{\mathbb{Z}}$,
given as the solution to the ACD equation \eqref{eq: EACD model}, and state
explicit conditions for stationarity and geometric ergodicity of $\{x_{i}\}$
as well as for power-law tails of $x$ with index $\kappa$. The range of the
values $\kappa$ will be crucial for our asymptotic\ theory.

The results are initially stated for general positive i.i.d. distributed
innovations $\{\varepsilon_{i}\}$.

\begin{lemma}
[ACD properties]\label{lem: tail index ACD} Consider $\left\{  x_{i}\right\}
$ given by \eqref{eq: EACD model} with a strictly positive i.i.d.
sequence\ $\{\varepsilon_{i}\}$ with density $f_{\varepsilon}$, and for which
$\mathbb{E}[\varepsilon]=1$ and $s^{2}=\mathbb{E}[\varepsilon^{2}]<\infty$.
Then $\left\{  x_{i}\right\}  $ is geometrically ergodic and has a stationary
representation for $\alpha\in(0,a_{u})$, $a_{u}=\exp\left(  -\mathbb{E}%
[\ln(\varepsilon)]\right)  >1$.$\,$Moreover, if the unique positive solution
$\kappa=\kappa(\alpha)>0$ to the equation $\mathbb{E}[(\alpha\varepsilon
)^{\kappa}]=1$ exists, then $\mathbb{P}(x>z)\sim c_{\kappa}\,z^{-\kappa}\,,$
$z\rightarrow\infty\,,$ for some positive constant $c_{\kappa}$ given in
\eqref{eq:constn}. In particular, we have%
\[
\left\{
\begin{array}
[c]{ll}%
2<\kappa<\infty\,, & \text{for }\alpha\in(0,1/s)\\
\kappa=2\,, & \text{for }\alpha=1/s\\
1<\kappa<2\,, & \text{for }\alpha\in(1/s,1)\\
\kappa=1\,, & \text{for }\alpha=1\\
0<\kappa<1\,, & \text{for }\alpha\in\left(  1,a_{u}\right)
\end{array}
\right.
\]

\end{lemma}

The results in Lemma \ref{lem: tail index ACD} complement existing results on
ARCH\ processes; see e.g. Embrechts, Kl\"{u}ppelberg and Mikosch (1997),
Buraczewski, Damek and Mikosch (2016), and allow in particular one to assess
the existence of moments of ACD\ processes. Thus, we find that for $\alpha<1$
the mean is finite, $\mathbb{E}[x]<\infty$, while the variance $\mathbb{V}[x]$
is finite for $\alpha$ in the smaller region $(0,1/s)$, where $s^{2}%
=\mathbb{E}[\varepsilon^{2}]$. For $1<\alpha<a_{u}$, while $\{x_{i}\}$ is a
strictly stationary and geometrically ergodic sequence, only fractional
moments (of order less than one) of $x$ are finite.

Next, we consider the benchmark model where $\varepsilon$ is exponentially
distributed (EACD).

\begin{lemma}
[EACD Properties]\label{lem: tail index} Consider $\left\{  x_{i}\right\}  $
given by \eqref{eq: EACD model} with an i.i.d. sequence\ $\{\varepsilon_{i}\}$
exponentially distributed with $\mathbb{E}[\varepsilon]=1$. The equation
\eqref{eq: EACD model} has a strictly stationary geometrically ergodic
solution $\{x_{i}\}$ if and only if\ $\alpha\in\lbrack0,a_{u})$, with
$a_{u}=\exp(\gamma)\simeq1.8$, where $\gamma$ is Euler's constant. The
remaining results in Lemma \ref{lem: tail index ACD} hold with $s^{2}%
=\mathbb{E}[\varepsilon^{2}]=2$; with $\Gamma$ denoting the Gamma function,
the equation $\mathbb{E}[(\alpha\varepsilon)^{\kappa}]=1$ has in this case a
unique implicit solution given by
\begin{equation}
\alpha=[\Gamma(\kappa+1)]^{-1/\kappa}\text{.} \label{eq: tail solution}%
\end{equation}

\end{lemma}

In particular, we observe the surprisingly simple explicit relationship
between $\alpha$ and $\kappa=\kappa\left(  \alpha\right)  $ in
\eqref{eq: tail solution} which comes from the properties of the exponential
distribution. {Such a simple relationship does not exist for general
distributions of $\varepsilon$ and more general functional forms of $\psi_{i}%
$.}

\subsection{Asymptotics for the ACD\ counting
process\label{sec: N convergence}}

In Lemma \ref{lem: N divided by t conv} below we collect some novel
asymptotic\ results for the counting process $N_{T}$, $T\geq0$, which are
needed for the asymptotic\ analysis of the QMLE of the ACD process.

Our results are general and of independent interest, in particular as the
dependence of the durations sequence\ is an uncommon condition in the
literature on renewal theory; there it is typically assumed that the durations
are i.i.d. or at most $m$-dependent (e.g. finite moving average); see e.g. Gut
(2009), Janson (1983). Moreover, and also new with respect to existing theory,
we present results for the convergence of the counting process $N_{T}$ when
durations have a tail index $\kappa<1$.

Recall initially that $N_{T}$ is defined in terms of the dependent sequence
$\left\{  x_{i}\right\}  $, cf. \eqref{eq: random summation index}, with
$x_{i}$ defined in (\ref{eq: EACD model}). As in Lemma
\ref{lem: tail index ACD} we consider here the general case of positive i.i.d.
innovations $\left\{  \varepsilon_{i}\right\}  $ with unit mean. The following
result provides convergence\ rates for $N_{T}$ as $T\rightarrow\infty$.

\begin{lemma}
\label{lem: N divided by t conv} Consider a strictly stationary geometrically
ergodic positive solution $\{x_{i}\}$ to \eqref{eq: EACD model} with tail
index $\kappa>0$ and $\{\varepsilon_{i}\}$ an i.i.d. sequence\ with
$\varepsilon>0$, $\mathbb{E}[\varepsilon]=1$. Then the following results hold
for the counting process $N_{T}$, $T\geq0$, defined in \eqref{eq: random summation index}.

\begin{enumerate}
\item[\textrm{(i)}] For $\kappa>1$,
\[
N_{T}/T\overset{\mathrm{a.s.}}{\rightarrow}1/\mu\,,\text{ \ \ where }%
\mu=\mathbb{E}[x]<\infty\,.
\]
\noindent

\item[\textrm{(ii)}] For $\kappa>2$, the CLT holds:
\[
T^{1/2}(N_{T}/T-1/\mu)\overset{d}{\rightarrow}N(0,\sigma^{2}/\mu^{3})\,,
\]
\noindent where $\sigma^{2}=\mathbb{E[}(1+T_{\infty})^{2}-T_{\infty}%
^{2}]\mathbb{V[}x]$ and $T_{\infty}=\sum_{i=1}^{\infty}\alpha^{i}(\prod
_{j=1}^{i}\varepsilon_{j})$. For $\kappa=2$ the CLT\ holds with normalization
$c\,\sqrt{T\,\ln T}$ for some positive constant $c$, and a standard normal
limit distribution.

\item[\textrm{(iii)}] For $1<\kappa<2,$%
\[
T^{\left(  \kappa-1\right)  /\kappa}(N_{T}/T-1/\mu)\overset{d}{\rightarrow
}\gamma_{\kappa}=\left(  c_{\kappa}\mathbb{E}\left[  (1+T_{\infty})^{\kappa
}-T_{\infty}^{\kappa}\right]  /\mu\right)  ^{1/\kappa}\eta_{\kappa}\,,
\]
\noindent where $\eta_{\kappa}$ is a totally skewed to the right $\kappa
$-stable random variable whose characteristic function\ is given in
\eqref{eq:chfct vs2}, and $c_{\kappa}$ is defined in
\eqref{eq:constn}%
.

\item[\textrm{(iv)}] For $0<\kappa<1$, $N_{T}/T\overset{\mathrm{a.s.}%
}{\rightarrow}0$ and
\begin{equation}
\dfrac{N_{T}}{T^{\kappa}}\overset{d}{\rightarrow}\lambda_{\kappa}=\left(
c_{\kappa}\,\mathbb{E}[(1+T_{\infty})^{\kappa}-T_{\infty}^{\kappa}]\right)
^{-1}\eta_{\kappa}^{-\kappa}, \label{eq lambdakappa case iv}%
\end{equation}
\noindent where $\eta_{\kappa}$ is a totally skewed to the right $\kappa
$-stable random variable whose characteristic function\ is given in
\eqref{eq:chfct vs2}, and $c_{\kappa}$ is defined in
\eqref{eq:constn}%
.
\end{enumerate}
\end{lemma}

It is worth noticing that for all cases (i)--(iv), $N_{T}\rightarrow\infty$
a.s. as a consequence\ of the ergodic theorem. However, the convergence\ rates
are quite distinct, depending on $\kappa$. Thus, $N_{T}/T\rightarrow1/\mu$
a.s. for $\kappa>1$, while, for $\kappa<1$, $N_{T}/T^{\kappa}$ converges in
distribution\ to the positive random variable $\lambda_{\kappa}$, and in
particular, $N_{T}/T\overset{p}{\rightarrow}0$. For $\kappa>2$, $N_{T}%
/T-1/\mu$ satisfies the CLT\ with standard $\sqrt{T}$-rate, while for
$1<\kappa<2$, the rate $T^{(\kappa-1)/\kappa}$ gets slower as $\kappa$ gets
closer to 1. We also note that the $\kappa$-stable limiting random variable
$\eta_{\kappa}$ has power-law tail{ with index $\kappa$.} Importantly, for the
novel result on the distributional convergence of $N_{T}/T^{\kappa}$ for
$\kappa<1$, the limiting variable $\lambda_{\kappa}$ has exponentially
decaying tails; cf. Theorem 2.5.2 in Zolotarev (1986).

\section{Asymptotic Theory for the QMLE}

\label{sec asymptotics} In this section we derive the asymptotic properties of
the (Q)MLE\ $\hat{\theta}_{T}=\arg\max_{\theta}L_{T}\left(  \theta\right)  $,
with $L_{T}\left(  \theta\right)  $ defined in (\ref{eq: EACD likelihood}).
Note that, as is common practice, $L_{T}\left(  \theta\right)  $ is defined
without the additional term corresponding to the fact that no events are
observed in the end-period $(t_{N_{T}},T]$. We show in Appendix
\ref{Appendix B} that this term has no influence on the asymptotic results.

We start in Section \ref{sec the score} by discussing the behavior of the
score and information, which is key to the asymptotic analysis. Then, in
Section~\ref{sec: QMLE}, we present the main results on the asymptotic
behavior of $\hat{\theta}_{T}$.

\subsection{Convergence of the score and information\label{sec the score}}

With the likelihood function $L_{T}(\theta)$ as given in
(\ref{eq: EACD likelihood}), the corresponding score and information
functions, evaluated at the true value $\theta=\theta_{0}$, are given by%
\begin{align}
S_{T}  &  =\left.  \tfrac{\partial L_{T}(\theta)}{\partial\theta}\right\vert
_{\theta=\theta_{0}}=\sum_{i=1}^{N_{T}}\xi_{i}\,,\quad\xi_{i}=(\varepsilon
_{i}-1)\,{\mathbf{v}}_{i}\,,\qquad{\mathbf{v}}_{i}=(1,x_{i-1})^{\prime}%
/\psi_{i}\,,\label{eq: score def}\\
I_{T}  &  =\left.  -\tfrac{\partial^{2}L_{T}(\theta)}{\partial\theta
\partial\theta^{\prime}}\right\vert _{\theta=\theta_{0}}=\sum_{i=1}^{N_{T}%
}\zeta_{i}\,,\qquad\zeta_{i}=(2\varepsilon_{i}-1)\,{\mathbf{v}}_{i}%
\,{\mathbf{v}}_{i}^{\prime}\,, \label{eq: info def}%
\end{align}
where $\psi_{i}=\psi_{i}(\theta_{0})$. In what follows, we always assume that
the conditions of Lemma~\ref{lem: tail index ACD} are satisfied. In
particular, \eqref{eq: EACD model} has a strictly stationary geometrically{
ergodic} solution $\{x_{i}\}$ with tail index $\kappa>0$.

Consider first the case $\kappa>1$, where we have the following result on the
large sample behavior of $S_{T}$ and $I_{T}$ at standard rates of convergence.

\begin{lemma}
\label{lem: score}Assume that for $\theta_{0}=\left(  \omega_{0},\alpha
_{0}\right)  ^{\prime}$, $\omega_{0}>0$ and $\alpha_{0}>0$ such that $\left\{
x_{i}\right\}  $ in \eqref{eq: EACD model} is stationary and ergodic, with
$\kappa>1$. With $\tau=\mathbb{V}[\varepsilon]$ and $\Omega=\mathbb{E}%
[{\mathbf{v}}_{1}{\mathbf{v}}_{1}^{\prime}]$ we have%
\[
T^{-1/2}S_{T}\overset{d}{\rightarrow}\left(  \tau\Omega/\mu\right)
^{1/2}{\mathbf{Z}}\text{\ and \ }T^{-1}I_{T}\overset{\mathrm{a.s.}%
}{\rightarrow}\Omega/\mu,
\]
where $\mathbf{Z}$ is a bivariate standard Gaussian vector. Moreover,
$N_{T}^{-1}I_{T}\overset{\mathrm{a.s.}}{\rightarrow}\Omega$.
\end{lemma}

Next turn to the case \emph{$\kappa<1$ }such that $\mathbb{E}[x]=\infty$. As
shown in the next, the score and information converge at slower rates than
usual. More specifically, turning to the information, it follows by Lemma
\ref{lem: N divided by t conv} that $N_{T}/T^{\kappa}\overset{d}{\rightarrow
}\lambda_{\kappa}$ and (see the proof of Lemma \ref{lem: score for k < 1}
below) $N_{T}^{-1}I_{T}\overset{\mathrm{a.s.}}{\rightarrow}\Omega$. Hence,%
\begin{equation}
T^{-\kappa}I_{T}\overset{d}{\rightarrow}\lambda_{\kappa}\Omega\text{.}
\label{eq. info limit vs2}%
\end{equation}
That is, the rate of convergence is indeed slower than standard when
$\kappa<1$, and the observed information is random in the limit due to the
random variable $\lambda_{\kappa}$. Similarly, non-standard convergence rates
as a function of $\kappa$ also apply to the score as we state the following
lemma for the EACD.

\begin{lemma}
\label{lem: score for k < 1} Assume that for $\theta_{0}=\left(  \omega
_{0},\alpha_{0}\right)  ^{\prime}$, $\omega_{0}>0$ and $\alpha_{0}>0$, such
that $\left\{  x_{i}\right\}  $ in \eqref{eq: EACD model} is stationary and
ergodic with $\kappa<1$, and $\varepsilon_{i}$ exponentially distributed with
$\mathbb{E}[\varepsilon]=1$. With $\Omega$ defined in Lemma \ref{lem: score}
we have%
\[
(T^{-\kappa/2}S_{T},T^{-\kappa}I_{T})\overset{d}{\rightarrow}\big((\lambda
_{\kappa}\Omega)^{1/2}\mathbf{Z},\lambda_{\kappa}\Omega\big)\text{ ,}%
\]
where{ }$\mathbf{Z}$ {a bivariate standard Gaussian vector, }independent of
$\lambda_{\kappa}$ defined in \eqref{eq lambdakappa case iv}{. }Moreover,
$N_{T}^{-1}I_{T}\overset{\mathrm{a.s.}}{\rightarrow}\Omega$.
\end{lemma}

\subsection{Limit theorems for the QMLE\label{sec: QMLE}}

We are now in the position to state the asymptotic distribution of the QMLE
$\hat{\theta}_{T}$ of $\theta$. As for the score, the limit behavior of the
QMLE\ depends on the tail behavior of the durations $\left\{  x_{i}\right\}
$. As mentioned, the influence of the right power-law tail of $x$ on the QMLE
is in contrast to QMLE theory for ARCH and GARCH processes where the shape of
the unconditional distribution does not matter. We show here that for
ACD\ processes the power-law tails determine the limiting distribution\ of the
QMLE $\hat{\theta}_{T}$ as well as the rate of convergence. This result
appears surprising, given that, apart from the random summation index, the ACD
(log-)likelihood function is identical to the ARCH Gaussian likelihood function.

Specifically, while $\sqrt{T}$-asymptotic normality holds when the tail index
$\kappa$ is above one, when $\kappa<1$, the speed of convergence and the
limiting distribution are non-standard. In particular, for the case $\kappa>1$
the following result holds.

\begin{theorem}
\label{thm: new QMLE}Under the assumptions of Lemma \ref{lem: score}, with
probability tending to one, there exists a local maximum $\hat{\theta}_{T}$ of
$L_{T}\left(  \theta\right)  $ which satisfies $\hat{\theta}_{T}\overset
{p}{\rightarrow}\theta_{0}$ and $\left.  \partial L_{T}\left(  \theta\right)
/\partial\theta\right\vert _{\theta=\hat{\theta}_{T}}=0$. Moreover,%
\begin{equation}
T^{1/2}(\hat{\theta}_{T}-\theta_{0})\overset{d}{\rightarrow}\left(  \Omega
/\mu\right)  ^{-1/2}\tau^{1/2}{\mathbf{Z}}, \label{eq: asymp normality 1}%
\end{equation}
with $\mathbf{Z}$ a bivariate standard Gaussian vector and $\mu=\mathbb{E}[x]$.
\end{theorem}

Theorem \ref{thm: new QMLE}, which is based on combining classic likelihood
expansions with the results for a random summation index $N_{T}$ in Section
\ref{sec: prelim}, shows that asymptotic normality at the $\sqrt{T}$-rate
holds even if the durations have infinite variance, $\mathbb{E}[x^{2}]=\infty
$. However, it can be shown that the quality of the asymptotic approximation
deteriorates as the tail index $\kappa=\kappa(\alpha_{0})$ approaches one.
This reflects the fact that the asymptotic\ results for the QMLE when
$\kappa>1$ are essentially derived by replacing the random indices $N_{T}$ in
the likelihood function\ (and its derivatives)\ by the deterministic
function\ $T/\mu$;\ this replacement, however, happens with a much larger
error term for $\kappa\in\left(  1,2\right)  $ than in the finite variance
case ($\kappa>2$), due to slow convergence\ rates of $N_{T}/T-1/\mu$ (cf.
Lemma \ref{lem: N divided by t conv}) and the widespread limit distribution.

\begin{figure}[t]
\begin{center}
\includegraphics[
trim=0.000000in 0.009925in 0.002049in -0.009925in,
height=7.7195cm,
width=16.1315cm
]{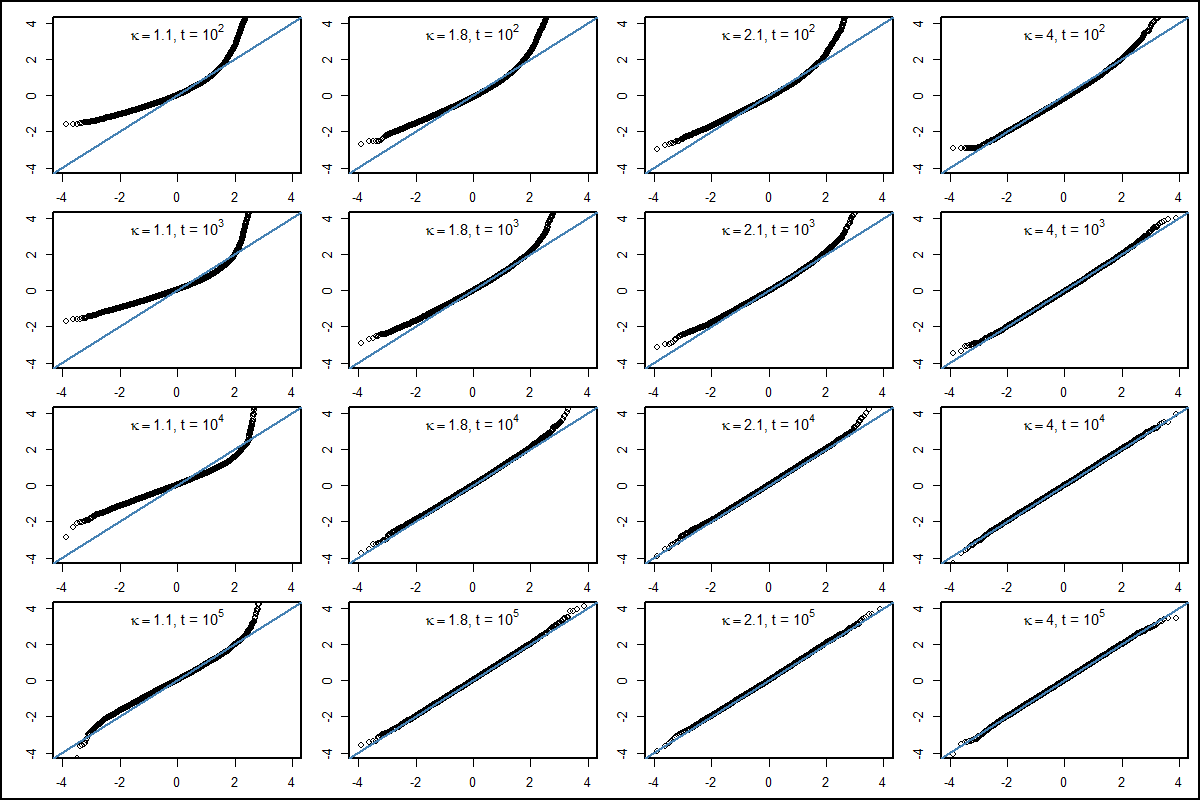}
\end{center}
\caption{Finite mean case, $\kappa>1$. Q-Q plots of $T^{1/2}(\hat{\alpha}%
_{T}-\alpha_{0})/\sigma_{\alpha}$, with $\sigma_{\alpha}$ the asymptotic
variance of $\hat{\alpha}_{T}$, against the $N\!\!\left(  0,1\right)  $
distribution for different values of $T$ (rows) and $\kappa_{0}=\kappa
(\alpha_{0})$ (columns), finite mean case ($\kappa>1$). $M=10^{4}$ Monte Carlo
replications. }%
\label{Figure kappa greater than one}%
\end{figure}

As an explanation to the fact that while the rate of convergence is standard
$\sqrt{T}$ for all $\kappa>1$, the convergence to the Gaussian limit for
$\kappa\in(1,2)$ slows down when compared to the (finite variance) case
$\kappa>2$, consider here the score $S_{T}$. By {Lemma
\ref{lem: N divided by t conv} (iii),}%
\[
T^{-1/2}S_{T}=[T^{\left(  1-\kappa\right)  /\kappa}\hat{\gamma}_{\kappa
}/(2{\sqrt{\mu})}+1/\sqrt{\mu}+o_{p}\left(  1\right)  ]\hat{Z}_{T},
\]
where $\hat{\gamma}_{\kappa}=T^{\left(  \kappa-1\right)  /\kappa}\left(
N_{T}/T-1/\mu\right)  \rightarrow_{d}\gamma_{\kappa}$ (a $\kappa$-stable
random variable) and $\hat{Z}_{T}=N_{T}^{-1/2}S_{T}\rightarrow_{d}\left(
\tau\Omega\right)  ^{1/2}\,\boldsymbol{Z}$. Additionally, $\gamma_{\kappa}$ is
non-standard distributed with {a power-law tail with index $\kappa$, and is
more widespread as }$\kappa$ diminishes. Thus, as $\kappa$ approaches one,
$T^{\left(  1-\kappa\right)  /\kappa}\hat{\gamma}_{\kappa}/(2{\sqrt{\mu})}$
converges to zero at a slower speed, and the convergence (in distribution) of
$T^{-1/2}S_{T}$ to the Gaussian distribution slows down{. This is in contrast}
to the case $\kappa>2$, where by Lemma~ \ref{lem: N divided by t conv} (ii),
$T^{-1/2}S_{T}=[T^{-1/2}\hat{\eta_{T}}/{(2\sqrt{\mu})}+1/\sqrt{\mu}%
+o_{p}\left(  1\right)  ]\hat{Z}_{T}$, with $\hat{\eta_{T}}=T^{1/2}%
(N_{T}/T-1/\mu)$ asymptotically Gaussian. In particular, the rate is
independent of $\kappa$ in this case.

We illustrate this in Figure \ref{Figure kappa greater than one}, where we
report Q-Q plots of $T^{1/2}(\hat{\alpha}_{T}-\alpha_{0})$ against the
Gaussian distribution when the data follows an EACD process with
$\mathbb{E}[x]=1$, for different values of $T$ and $\kappa$. The figure
clearly shows how the tail index of the durations influences the quality of
the Gaussian approximation in finite time intervals. It can also be seen that
as $\kappa$ gets closer to one, the asymptotic approximation requires larger
values of $T$ to be accurate.

For $\kappa<1$, as previously emphasized, $\hat{\theta}_{T}-\theta_{0}$ is not
asymptotically Gaussian distributed.

\begin{theorem}
\label{Theorem EACD MLE kappa <1}Under the assumptions of Lemma
\ref{lem: score for k < 1}, with probability tending to one, there exists a
local maximum $\hat{\theta}_{T}$ of $L_{T}\left(  \theta\right)  $ which
satisfies $\hat{\theta}_{T}\overset{p}{\rightarrow}\theta_{0}$ and $\left.
\partial L_{T}\left(  \theta\right)  /\partial\theta\right\vert _{\theta
=\hat{\theta}_{T}}=0$. Moreover,
\begin{equation}
T^{\kappa/2}(\hat{\theta}_{T}-\theta_{0})\overset{d}{\rightarrow}\left(
\lambda_{\kappa}\Omega\right)  ^{-1/2}{\mathbf{Z}},
\label{eq asy for thetahat in infinite mean}%
\end{equation}
with $\mathbf{Z}$ a bivariate standard Gaussian vector, independent of
$\lambda_{\kappa}$ defined in \eqref{eq lambdakappa case iv}.
\end{theorem}

Thus for $\kappa<1$, the estimators are asymptotically mixed Gaussian;
moreover the rate of convergence $T^{\kappa/2}$ is lower than the standard
$T^{1/2}$ rate and depends on the value of $\kappa$. The non-Gaussianity in
(\ref{eq asy for thetahat in infinite mean}) is clearly illustrated in the
upper panel of Figure \ref{klessthanone}, which reports Q-Q plots of
$T^{\kappa/2}(\hat{\alpha}_{T}-\alpha_{0})$ against a zero-mean Gaussian
distribution for $\kappa=0.5$ and different values of $T$ ($\omega_{0}$ is
selected such that the median of $x_{i}$ is about one).

\section{Discussion and implications for inference}

\label{sec discussion}

In the previous section we have shown that, for the case of a finite mean of
the durations, the (Q)MLE\ is indeed asymptotically normal at the standard
$\sqrt{T}$-rate while, for the case of infinite mean, the limiting
distribution is a mixture and convergence attains at a lower rate.

\begin{figure}[t]
\begin{center}
\includegraphics[
trim=0.000000in 0.005504in 0.002033in -0.005504in,
height=5.7195cm,
width=16.1315cm
]{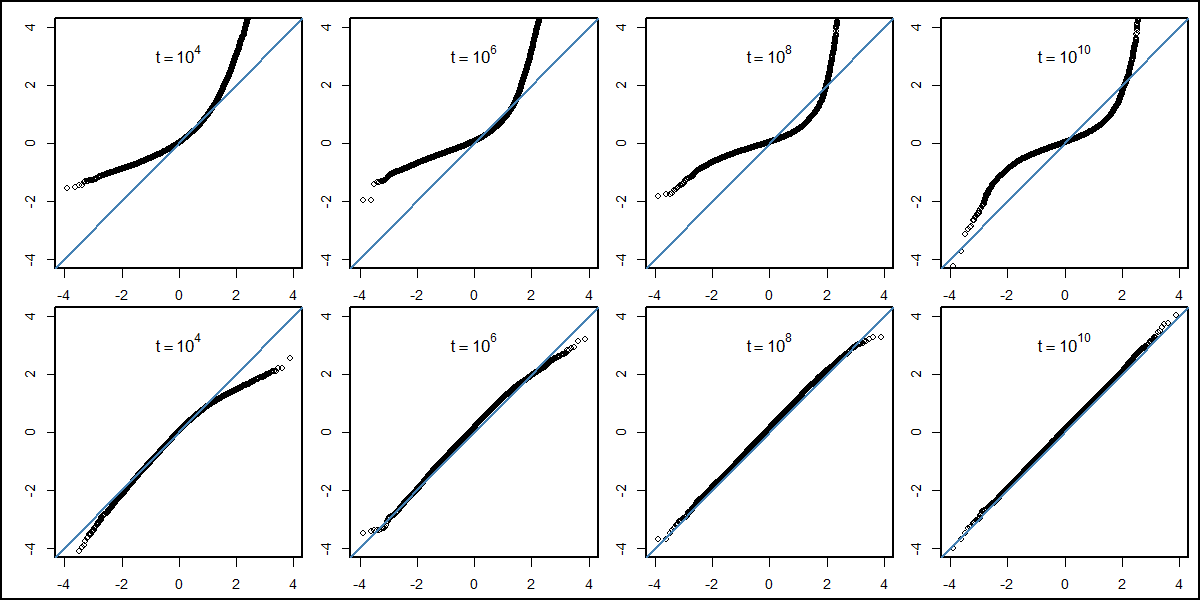}
\end{center}
\caption{Infinite mean case, $\kappa=0.5$. Q-Q plots against the $N\!\!\left(
0,1\right)  $ distribution. Upper panel:\ $T^{\kappa/2}(\hat{\alpha}%
_{T}-\alpha_{0})$ (normalized by empirical variance). Lower panel: t-ratios.
$M=10^{4}$ Monte Carlo replications. }%
\label{klessthanone}%
\end{figure}

In terms of inference, from the Theorems \ref{thm: new QMLE} and
\ref{Theorem EACD MLE kappa <1} we can derive the following new result, which
shows that $t$-ratios (or quasi likelihood ratio statistics) are
asymptotically standard Gaussian ($\chi^{2}$) distributed, irrespectively of
the tail index of the durations $\kappa$ being above or below unity. That is,
while $\kappa$ affects the distributional theory for of (Q)MLE, asymptotic
inference based on $t$ tests (or likelihood ratio tests)\ is standard, and
asymptotic validity holds irrespective of the tail index of the marginal
distribution of the durations $\{x_{i}\}$.

\begin{corollary}
\label{Cor t}For $\kappa>1$, and under the assumptions of Theorem
\ref{thm: new QMLE} and, for $\kappa<1$ and under the assumptions of Theorem
\ref{Theorem EACD MLE kappa <1}, the $t$ ratio $t_{n}=\operatorname*{se}%
_{T}^{-1}(\hat{\alpha}_{T}-\alpha_{0})$, where $\operatorname*{se}_{T}^{2}$ is
the entry of $I_{T}(\hat{\theta}_{T})=\left.  -\tfrac{\partial^{2}L_{T}%
(\theta)}{\partial\theta\partial\theta^{\prime}}\right\vert _{\theta
=\hat{\theta}_{T}}$ corresponding to $\alpha$, is standard normal as
$T\rightarrow\infty$.
\end{corollary}

Convergence of the $t$ ratios to the Normal distribution is illustrated in the
lower panel of Figure \ref{klessthanone}, where Q-Q plots of $t_{n}$ against
the $N\!\!\left(  0,1\right)  $ distribution are reported for increasing
sample sizes and for $\kappa<1$. The figure clearly shows that extremely large
observation periods are required for the normal asymptotic approximation to be
accurate. This implies that in empirical applications, and differently from
inference in ARCH models, inspection of the tails of the marginal distribution
of the data is a key step to be taken prior to any empirical analysis.

Finally, we note that, in terms of theory, the result in Corollary \ref{Cor t}
for $\kappa<1$ is similar to the mixed Gaussian limit results, as employed
e.g. in the cointegration analysis of non-stationary variables; see Johansen
(1991) and Phillips (1991).

\section{Conclusions}

Our new results demonstrate the sensitivity of the limiting distribution of
the QMLE\ in ACD\ models to the tail behaviour, or equivalently finiteness of
moments, of the durations. This clearly contrasts the previous asymptotic
results which, by treating the number of durations as deterministic and hence
referring to ARCH\ asymptotic theory, does not depend on the finiteness of
moments, nor on the tail behaviour. Stated differently, sampling over a fixed
period of time (hence implying a random number of events) leads to new
non-standard theory, while sampling over a fixed number of events (hence
implying a random length of observation period) leads to standard\ theory from
ARCH models.

All results can be generalized to more general ACD\ models, in particular to
the much applied ACD model where $\psi_{i}=\omega+\alpha x_{i-1}+\beta
\psi_{i-1}$, that is, the ACD\ analogue of the GARCH(1,1), as well as its
extensions. We have refrained from doing so here to keep the presentation
simple, and thereby focus on the main new insights.

Finally, it is worth noticing that our findings and arguments, are not
specific to the models for time series of durations. Indeed, they apply to any
econometric method where the number of observations needs being treated as
random. For example, asymptotic theory for daily realized volatility, see Li,
Mykland, Renault, Zhang and Zheng (2013), treats summations such as
$\sum_{i=1}^{N_{T}}(p_{t_{i}}-p_{t_{i-1}})^{2}$, where $p_{t}$ is the
(log-)price at time $t$; since the number of trades within a day, $N_{T}$, is
random, our results could be applied to cases where $x_{i}=t_{i}-t_{i-1}$ have
heavy tails.

\section*{Acknowledgements}

We are grateful to Federico Bandi, Marcelo Fernandes, Nikolaus Hautsch and
Marcelo Medeiros for comments and discussions. We also thank participants at
the SoFiE 2022 conference (U Cambridge), the Aarhus Workshop in Econometrics
(Aarhus U), the 3rd High Voltage Econometrics meeting and the
Bologna/Rome-Waseda Time Series Workshop. We also thank Roberto Ren\`{o} for
providing the SPY duration data. A. Rahbek and G. Cavaliere gratefully
acknowledge support from the Danish Council for Independent Research (DSF
Grant 015-00028B). Part of G. Cavaliere's research was supported by the
Italian Ministry of University and Research (PRIN 2020 Grant 2020B2AKFW).
Thomas Mikosch's research is partially supported by Danmarks Frie
Forskningsfond Grant No 9040-00086B.

\section*{References}

\noindent\textsc{Allen, D., Felix, C., McAleer, M. and Peiris, S.} (2008)
Finite sample properties of the QMLE for the log-ACD model: Application to
Australian Stocks. \emph{Journal of Econometrics,} 147:163--185.

\smallskip\noindent\textsc{Aquilina, M., Budish, E. and O'Neill, P.} (2022)
Quantifying the high-frequency trading \textquotedblleft Arms
Race\textquotedblright. Quarterly Journal of Economics 137:493--564,

\smallskip\noindent\textsc{Billingsley, P.}\ (1999) \emph{Convergence of
Probability Measures.} Wiley, NY.

\smallskip\noindent\textsc{Bingham, N., Goldie, C. and Teugels, J.} \ (1987)
\emph{Regular Variation.} Cambridge University Press, Cambridge UK.

\smallskip\noindent\textsc{Buraczewski, D., Damek, E. and Mikosch, T.}\ (2016)
\emph{Stochastic Models with Power-Law Tails.} Springer, NY.

\smallskip\noindent\textsc{Cavaliere, G., Lu, Y. , Rahbek, A. and
St\ae rk-\O stergaard, J.} (2022) Bootstrap inference for Hawkes and general
point processes. \emph{Journal of Econometrics, }in press.

\smallskip\noindent\textsc{Daley, D.J. and Vere-Jones, D.} (2008) \emph{An
Introduction to the Theory of Point Processes}, Volume II:\ General Theory and
Structure. Springer, Berlin.

\smallskip\noindent\textsc{Embrechts, P., Kl\"{u}ppelberg, C. and
Mikosch,\ M.}\ (1997) \emph{Modelling Extremal Events: For Insurance and
Finance.} Springer, Heidelberg.

\smallskip\noindent\textsc{Embrechts, P., Liniger, T. and Lin, L.} (2011)
Multivariate Hawkes processes: an application to financial data. \emph{Journal
of Applied Probability}, 48:367--378.

\smallskip\noindent\textsc{Engle, R.F. }(2000) The econometrics of
ultra-high-frequency data. \emph{Econometrica}, 68:1--22.

\smallskip\noindent\textsc{Engle, R.F. and Russell, J.R.}\ (1998)
Autoregressive conditional duration: a new model for irregularly spaced
transaction data. \emph{Econometrica}, 66:1127--1162.

\smallskip\noindent\textsc{Fernandes, M. and Grammig, J. }(2006) A family of
autoregressive conditional duration models. \emph{Journal of Econometrics,
}130: 1--23.

\smallskip\noindent\textsc{Fernandes, M., Medeiros, M.C., A. and Veiga }(2016)
The (semi-)parametric functional coefficient autoregressive conditional
duration model. \emph{Econometric Reviews }35:1221--1250.

\smallskip\noindent\textsc{Guivarc'h, Y. and Le Page, E.}\ (2008) On spectral
properties of a family of transfer operators and convergence\ to stable laws
for affine random walks. \emph{Ergodic Theory and Dynamical Systems}, 28:423--446.

\smallskip\noindent\textsc{Gut, A.}\ (2009) \emph{Stopped Random Walks: Limit
Theorems and Applications}. Springer, NY.

\smallskip\noindent\textsc{Hamilton, J.D. and \`{O}. Jord\`{a} }(2002): A
Model of the Federal Funds Rate Target. \emph{Journal of Political Economy}, 110:1135--1167.

\smallskip\noindent\textsc{Hautsch, N.}\ (2012) \emph{Econometrics of
Financial High-Frequency Data.} Springer, Berlin.

\smallskip\noindent\textsc{Janson, S.}\ (1983) Renewal theory for
$m$-dependent variables. \emph{Annals of Probability}, 11:558--568.

\smallskip\noindent\textsc{Jensen, S.T. and Rahbek, A.} (2004) Asymptotic
normality of the QMLE estimator of ARCH in the nonstationary case.
\emph{Econometrica, }72:641--646.

\smallskip\noindent\textsc{Krengel, U.} (1985) \emph{Ergodic Theorems.} De
Gryuter, Berlin.

\smallskip\noindent\textsc{Johansen, S.} (1991)\ Estimation and hypothesis
testing of cointegration vectors in Gaussian vector autoregressive models.
\emph{Econometrica}, 59:1551--1580.

\smallskip\noindent\textsc{Kristensen, D. and Rahbek, A.} (2010)
Likelihood-based inference for cointegration with nonlinear error-correction.
\emph{Journal of Econometrics, }158:78--94.

\smallskip\noindent\textsc{Lee, S. and Hansen, B.}\ (1994) Asymptotic theory
for the \textrm{GARCH}$(1,1)$\ quasi-maximum likelihood estimator.
\emph{Econometric Theory,} 10:29--52.

\smallskip\noindent\textsc{Li, Y., Mykland, P.A., Renault, E., Zhang, L., and
Zheng, X.} (2014) Realized volatility when sampling times are possibly
endogenous. \emph{Econometric Theory}, 30:580--605.

\smallskip\noindent\textsc{Pacurar, M.}\ (2008) Autoregressive conditional
duration models in finance: a survey of the theoretical and empirical
literature. \emph{Journal of Economic Surveys}, 22:711---751.

\smallskip\noindent\textsc{Phillips, P.C.B.} (1991)\ Optimal inference in
cointegrated systems, \emph{Econometrica}, 59:283--306.

\smallskip\noindent\textsc{Sin, C.} (2014) QMLE of a standard exponential
ACD\ model:\ Asymptotic distribution and residual correlation.\emph{ Annals of
Financial Economics}, Vol. 09: Issue 2.

\smallskip\noindent\textsc{Sweeting}, \ (1980) Uniform asymptotic normality of
the maximum likelihood estimator. \emph{Annals of Statistics}, 8:1375--1381.

\smallskip\noindent\textsc{Zolotarev, V.M.} (1986) \emph{One-Dimensional
Stable Distributions}. American Mathematical Society, Providence, Rhode Island.

\appendix

\section*{Appendix}

\section{Proofs}

\subsection*{Proof of Lemma~\ref{lem: tail index ACD}}

Observe that the ACD equation \eqref{eq: EACD model} can be formulated as a
stochastic recurrence equation (SRE):
\begin{equation}
x_{i}=A_{i}\,x_{i-1}+B_{i}\text{,}\qquad i\in{\mathbb{Z}}\,, \label{eq SRE}%
\end{equation}
with a sequence\ $(A_{i},B_{i})=(\omega\,,\alpha)\varepsilon_{i}$,
$i\in{\mathbb{Z}}$, of random vectors with i.i.d. positive $\{\varepsilon
_{i}\}$. Using the SRE\ representation, it follows that $\{x_{i}\}$ is
strictly stationary geometrically ergodic if and only if\ $\mathbb{E}%
[\ln(\alpha\varepsilon)]<0$ by Theorem 2.1.3 and Proposition 2.2.4 in
Buraczewski, Damek and Mikosch (2016), BDM henceforth. The power-law tail
behavior{ }$\mathbb{P}${$(x>z)\sim c_{\kappa}\,z^{-\kappa}$} follows from
Theorem 2.4.4 in BDM, and it holds that $c_{\kappa}$ is given by
\begin{equation}
c_{\kappa}=\dfrac{\mathbb{E}[(\omega+(\alpha\,\varepsilon)\,x)^{\kappa
}-((\alpha\,\varepsilon)x)^{\kappa}]}{\kappa\,\mathbb{E}[(\alpha
\varepsilon)^{\kappa}\,\ln(\alpha\varepsilon)]}\,. \label{eq:constn}%
\end{equation}
\hfill\hfill$\square$

\subsection*{Proof of Lemma \ref{lem: tail index}}

The results hold by Lemma \ref{lem: tail index ACD}, noting that for the
exponential case,
\begin{equation}
1=\mathbb{E}[(\alpha\varepsilon)^{\kappa}]=\alpha^{\kappa}\int_{0}^{\infty
}x^{\kappa}\exp(-x)\,dx=\alpha^{\kappa}\,\Gamma(\kappa+1)\,.
\label{eq: identity vs2}%
\end{equation}
\hfill$\square$

\subsection*{Proof of Lemma \ref{lem: N divided by t conv}}

\emph{Convergence a.s. for $\kappa>1$.} Since $\kappa>1$ we have
$\mu=\mathbb{E}[x]<\infty$. We follow the argument in Theorem 5.1 in Gut
(2009). Since $\{x_{i}\}$ is ergodic $T_{n}/n=\sum_{i=1}^{n}x_{i}%
/n\overset{\mathrm{a.s.}}{\rightarrow}\mu$, hence $\nu_{T}=${$N_{T}$}%
${+1}\overset{\mathrm{a.s.}}{\rightarrow}\infty$ and $T_{\nu_{T}}/\nu
_{T}\overset{\mathrm{a.s.}}{\rightarrow}\mu$. But $T<T_{\nu_{T}}\leq
T+x_{\nu_{T}}$ and
\[
0<T_{\nu_{T}}/\nu_{T}-T/\nu_{T}\leq x_{\nu_{T}}/\nu_{T}\overset{\mathrm{a.s.}%
}{\rightarrow}0\,,
\]
\noindent hence $\nu_{T}/T\overset{\mathrm{a.s.}}{\rightarrow}1/\mu$.
\bigskip\newline\emph{Convergence in distribution for $\kappa\geq2$.} We start
with $\kappa>2$. We have
\begin{equation}
T^{-1/2}(T_{N_{T}}-\mu N_{T})\leq T^{-1/2}(T-\mu\,N_{T})\leq T^{-1/2}%
(T_{N_{T}}-\mu N_{T})+T^{-1/2}x_{\nu_{T}}\,. \label{eq: k>2 conv}%
\end{equation}
\noindent First, we prove that $T^{-1/2}x_{\nu_{T}}\overset{p}{\rightarrow}0$.
For $M,\delta>0$ we have
\begin{align*}
\mathbb{P}(T^{-1/2}x_{\nu_{T}}>M)  &  \leq\mathbb{P}(T^{-1/2}x_{\nu_{T}%
}>M,|\nu_{T}/T-1/\mu|>\delta)+\mathbb{P}(T^{-1/2}x_{\nu_{T}}>M,|\nu
_{T}/T-1/\mu|\leq\delta)\\
&  =I_{1}+I_{2}\,.
\end{align*}
\noindent But $I_{1}\rightarrow0$ as $T\rightarrow\infty$ for every $\delta>0$
by virtue of the first part of the proof. On the other hand, by stationarity,
\[
I_{2}\leq\mathbb{P}\left(  \max_{T(1/\mu-\delta)\leq s\leq T(1/\mu+\delta
)}x_{s}>T^{1/2}\,M\right)  \leq\mathbb{P}\left(  T^{-1/2}\max_{s\leq3\delta
T}x_{s}>M\right)  \,.
\]
\noindent The right-hand side\ converges to zero since $T^{-1/\kappa}%
\max_{s\leq T}x_{s}$ converges in distribution\ to a Fr\'{e}chet distribution;
see BDM, Theorem 3.1.1.

In view of (\ref{eq: k>2 conv}) we thus proved that the distributional limits
of $T^{-1/2}(T_{N_{T}}-\mu N_{T})$ and $T^{-1/2}(T-\mu\,N_{T})$ coincide if
they exist. However, Theorem 3.3.1 in BDM yields $n^{-1/2}(T_{n}%
-\mu\,n)\overset{d}{\rightarrow}N(0,\sigma^{2})$ as $n\rightarrow\infty$ with
$\sigma^{2}=E[(1+T_{\infty})^{2}-T_{\infty}^{2}]V[x]$ and\textbf{ }$T_{\infty
}=\sum_{i=1}^{\infty}\alpha^{i}(\prod_{j=1}^{i}\varepsilon_{j})$.

{In what follows, we will frequently abuse notation: when sums are involved
and their index is not a natural number we understand these expressions as
taken at their integer parts.} Abusing notation, we then have as
$T\rightarrow\infty$,%
\[
T^{-1/2}\left(  T_{T/\mu}-T\right)  =\mu^{-1/2}\left(  T/\mu\right)
^{-1/2}\sum_{i=1}^{T/\mu}\left(  x_{i}-\mu\right)  \overset{d}{\rightarrow
}N\left(  0,\sigma^{2}/\mu\right)  \text{ }.
\]
\noindent Then the CLT\ for $T^{-1/2}(T_{N_{T}}-\mu N_{T})$ will follow if we
can prove that for every $M>0$,
\[
I=\mathbb{P}\left(  T^{-1/2}\left\vert (T_{N_{T}}-\mu N_{T})-(T_{T/\mu
}-T)\right\vert >M\right)  \rightarrow0\,.
\]
\noindent We apply an Anscombe argument; see Gut (2009). For given
$M,\delta>0$ we have
\[
I\leq\mathbb{P}(|N_{T}/T-1/\mu|>\delta)+\mathbb{P}\left(  T^{-1/2}\left\vert
(T_{N_{T}}-\mu N_{T})-(T_{T/\mu}-T)\right\vert >M\,,|N_{T}/T-1/\mu|\leq
\delta\right)  =I_{3}+I_{4}\,.
\]
\noindent As before, $I_{3}\rightarrow0$ as $T\rightarrow\infty$. On the other
hand, by stationarity,
\begin{align*}
I_{4}  &  \leq\mathbb{P}\left(  \max_{T(1/\mu-\delta)\leq s\leq T(1/\mu
+\delta)}\left\vert T_{s}-T_{T/\mu}-\mu(s-T/\mu)\right\vert >T^{1/2}M\right)
\\
&  \leq2\,\mathbb{P}\left(  \max_{u\leq\delta}\left\vert T_{u\,T}%
-\mu\,u\,T\right\vert >T^{1/2}\,M\right)  \text{ }.
\end{align*}
\noindent Since $\{x_{i}\}$ is geometrically ergodic we can apply the
functional CLT\ with Brownian limit and the continuous mapping theorem; see
Theorem 19.1 in Billingsley (1999). Then the right-hand side\ vanishes by
first letting $T\rightarrow\infty$ and then $\delta\rightarrow0$.

The case $\kappa=2$ is similar but we have to replace the normalization
$T^{1/2}$ by {$C(T\ln T)^{1/2}$ for a suitable constant $C>0$.} The proof of
$x_{\nu_{T}}/(T\ln T)^{1/2}\overset{p}{\rightarrow}0$ follows in the same way
since $n^{-1/2}\max_{t=1,\ldots,n}x_{t}$ converges in distribution\ to a
Fr\'{e}chet distribution; {see BDM, Theorem 3.1.1.} A functional CLT\ with
Brownian limit and normalization $(T\,\ln T)^{1/2}$ is given in Guivarc'h and
Le Page (2008).

\bigskip

\emph{Convergence in distribution for $1<\kappa<2$.} Similar to the case
of\textbf{ }$\kappa>2$\textbf{,} the starting point is the
inequalities\textbf{,}%
\begin{equation}
T^{-1/\kappa}(T_{N_{T}}-\mu N_{T})\leq T^{-1/\kappa}(T-\mu N_{T})\leq
T^{-1/\kappa}(T_{N_{T}}-\mu N_{T})+T^{-1/\kappa}x_{\nu_{T}}\text{.}
\label{eq: inequal}%
\end{equation}
We observe that for $M,\delta>0$, by stationarity,
\begin{align*}
\mathbb{P}(x_{\nu_{T}}>T^{1/\kappa}\,M)  &  \leq\mathbb{P}(|\nu_{T}%
/T-\mu|>\delta)+\mathbb{P}(T^{-1/\kappa}x_{\nu_{T}}>M\,,|\nu_{T}/T-\mu
|\leq\delta)\\
&  \leq o(1)+\mathbb{P}\left(  T^{-1/\kappa}\max_{s\leq3\delta T}%
x_{s}>M\right)  \,,
\end{align*}
\noindent and the right-hand side\ converges to zero by first letting
$T\rightarrow\infty$ and then $\delta\rightarrow0$. In the last step one uses
the Fr\'{e}chet convergence\ of $T^{-1/\kappa}\max_{i=1,\ldots,T}x_{i}$.

Next we observe that by BDM, Theorem 3.3.4,%
\[
(c_{\kappa}n)^{-1/\kappa}(T_{n}-\mu\,n)\overset{d}{\rightarrow}\left(
\mathbb{E}\left[  (1+T_{\infty})^{\kappa}-T_{\infty}^{\kappa}\right]  \right)
^{1/\kappa}\eta_{\kappa}\,,
\]
\noindent where $c_{\kappa}$ is the constant in \eqref{eq:constn}, $T_{\infty
}$ is defined in the lemma and $\eta_{\kappa}$ is $\kappa$-stable with
characteristic function\
\begin{equation}
\varphi_{\eta_{\kappa}}(s)=\exp\left(  -\int_{0}^{\infty}\left(  \mathrm{\exp
}\left(  isy\right)  -1-isy\,\mathbb{I}(1<\kappa<2)\right)  \kappa
\,y^{-\kappa-1}dy\right)  \,,\qquad s\in{\mathbb{R}}\,. \label{eq:chfct vs2}%
\end{equation}
It remains to show that for every $M>0$.
\[
J=\mathbb{P}\left(  T^{-1/\kappa}\left\vert (T_{N_{T}}-\mu N_{T})-(T_{T/\mu
}-T)\right\vert >M\right)  \rightarrow0\,.
\]
\noindent We have for every $\delta>0$,
\[
J\leq o(1)+\mathbb{P}\left(  T^{-1/\kappa}\left\vert (T_{N_{T}}-\mu
N_{T})-(T_{T/\mu}-T)\right\vert >M\,,|N_{T}/T-1/\mu|\leq\delta\right)
=o(1)+J_{1}\,.
\]
\noindent Abusing notation, we have
\[
J_{1}\leq\mathbb{P}\left(  \max_{T\left(  1/\mu-\delta\right)  \leq s\leq
T\left(  1/\mu+\delta\right)  }\left\vert T_{s}-T_{T/\mu}-\mu\left(
s-T/\mu\right)  \right\vert >T^{1/\kappa}M\right)  \leq2\mathbb{P}\left(
\max_{u\leq T\delta}\left\vert T_{u}-\mu u\right\vert >T^{1/\kappa}M\right)
\]
\noindent On one hand, we observe that for fixed $\widetilde{\varepsilon}>0$,
\[
\mathbb{P}\left(  \max_{u\leq T\delta}|T_{u}-\mu u|>T^{1/\kappa}M,\max_{s\leq
T\delta}\,x_{s}>T^{1/\kappa}\,\widetilde{\varepsilon}\right)  \leq\delta
T\,\mathbb{P}(x>\widetilde{\varepsilon}T^{1/\kappa})\sim\mathrm{const}\text{
}\delta\,\widetilde{\varepsilon}^{-\kappa}\,,\text{ }T\rightarrow\infty.
\]
\noindent The right-hand side\ converges to zero as $\delta\rightarrow0$. Next
we mimic the proof of Theorem 4.5.2 in BDM. Write
\[
X_{T}=\overline{X}_{T}+\underline{X}_{T},\text{ \ \ }\overline{X}_{T}%
=X_{T}f(T^{-1/\kappa}X_{T})\,,\qquad\overline{T}_{n}=\sum_{i=1}^{n}%
\overline{X}_{i}\,,
\]
\noindent with $f(x)\in\lbrack0,1]$ smooth, $\operatorname*{supp}%
f\subset\{x:|x|\leq\widetilde{\varepsilon}\}$, and $f(x)=1$ for $|x|\leq
\widetilde{\varepsilon}/2$. Then%
\[
\mathbb{P}\left(  \max_{u\leq\delta T}|T_{u}-\mu u|>T^{1/\kappa}%
\,M,\max_{s\leq\delta T}x_{s}\leq T^{1/\kappa}\,\widetilde{\varepsilon
}\right)  \leq\mathbb{P}\left(  \max_{u\leq\delta T}|\overline{T}%
_{u}-u\,\mathbb{E}[\overline{x}]|+\delta T\,\mathbb{E}[\underline
{x}]>T^{1/\kappa}M\right)
\]
\noindent By Karamata's theorem (see Bingham, Goldie and Teugels, 1987) for
large $T$,
\[
\delta\,T^{1-1/\kappa}\,\mathbb{E}[\underline{x}]\geq\mathrm{const}%
\,\delta\widetilde{\varepsilon}^{1-\kappa}\frac{\mathbb{E}[x/(\widetilde
{\varepsilon}T^{1/\kappa})\mathbb{I}(x>\widetilde{\varepsilon}T^{1/\kappa}%
)]}{\mathbb{P}(x>\widetilde{\varepsilon}T^{1/\kappa})}\sim\mathrm{const}%
\,\delta\widetilde{\varepsilon}^{1-\kappa}\rightarrow0\text{ as }%
\delta\rightarrow0\text{.}%
\]
\noindent Therefore it is suffices to show that the following quantity
vanishes by first letting $T\rightarrow\infty$ and then $\delta\rightarrow0$:
\[
Q=\mathbb{P}\left(  \max_{u\leq\delta T}|\overline{T}_{u}-u\,\mathbb{E}%
[\overline{x}]|>T^{1/\kappa}\,M\right)  \,.
\]
\noindent With $s(T)=\delta\,T^{1-\beta}$ for $\beta\in(0,1)$, {(Here we
assume without loss of generality that $s(T)$ is an integer).}
\begin{align*}
Q  &  \leq\mathbb{P}\left(  \max_{k=1,...,s\left(  T\right)  }|\overline
{T}_{kT^{\beta}}-kT^{\beta}\mathbb{E}[\overline{x}]|>T^{1/\kappa}M\right) \\
&  +\mathbb{P}\left(  \max_{k=1,..,s(T)}\max_{u\in\{(k-1)T^{\beta}%
+1,\ldots,kT^{\beta}\}}|(\overline{T}_{u}-\overline{T}_{(k-1)T^{\beta}%
})-(s-(k-1)T^{\beta})\mathbb{E}[\overline{x}]|>T^{1/\kappa}M\right) \\
&  =Q_{1}+Q_{2}\,,
\end{align*}
\noindent{ignoring the last incomplete block of indices as it does not
contribute to the asymptotic theory.} Observe that $T^{\beta-1/\kappa
}\mathbb{E}[\overline{x}]\rightarrow0$ as $T\rightarrow\infty$ for
$\beta<1/\kappa$. {Hence by stationarity and for large $T$,}
\begin{align*}
Q_{2}  &  \leq s(T)\,\mathbb{P}\left(  \max_{u\leq T^{\beta}}|\overline{T}%
_{u}-u\,\mathbb{E}[\overline{x}]|>T^{1/\kappa}\,M\right)  \leq s(T)\mathbb{P}%
(\overline{T}_{T^{\beta}}>T^{1/\kappa}M/2)\\
&  \leq s(T)\mathbb{P}(\overline{T}_{T^{\beta}}-\mathbb{E}[\overline
{T}_{T^{\beta}}]>T^{1/\kappa}M/3)\leq\mathrm{const}\,\delta T^{(1-(2/\kappa
))(1-\beta)}\mathbb{V}[T^{-\beta/\kappa}\overline{T}_{T^{\beta}}]\text{.}%
\end{align*}
\noindent By the calculations on p. 211 in BDM, the variance on the right-hand
side\ is bounded. Hence $Q_{2}\rightarrow0$.

Now we turn to $Q_{1}$. For $k\leq s(T)$ we have for $\lambda>0$,%
\[
\mathbb{P}(|{\overline{T}}_{kT^{\beta}}-kT^{\beta}\mathbb{E}[\overline
{x}]|>\lambda)\leq\lambda^{-2}\mathbb{V[}(kT^{\beta})^{-1/\kappa}\overline
{T}_{kT^{\beta}}]k^{2/\kappa}\,T^{2\beta/\kappa}\leq\mathrm{const}%
\,\lambda^{-2}k^{2/\kappa}T^{2\beta/\kappa}\,,
\]
\noindent where we again used the variance bounds on p. 211 in BDM. An
application of Theorem 10.2 in Billingsley (1999) yields
\[
Q_{1}\leq\mathrm{const}\,M^{-2}\,T^{-2/\kappa}\,T^{2\beta/\kappa}\,(\delta
T^{1-\beta})^{2/\kappa}=\mathrm{const}\,M^{-2}\,\delta^{2/\kappa}%
\rightarrow0\,,\qquad\delta\rightarrow0\,.
\]
\noindent This finishes the proof in the case $\kappa\in(1,2)$. \bigskip
\newline\emph{Convergence in distribution for $0<\kappa<1$.} Using Theorem
3.3.4 in BDM, for $z>0$%
\begin{align*}
\mathbb{P}(T^{-\kappa}N_{T}\leq z)  &  =\mathbb{P}(T_{zT^{\kappa}%
}>T)=1-\mathbb{P}(T_{zT^{\kappa}}\leq T)=1-\mathbb{P}\left(  (c_{\kappa
}zT^{\kappa})^{-1/\kappa}T_{zT^{\kappa}}\leq(c_{\kappa}\,z)^{-1/\kappa}\right)
\\
&  \rightarrow1-\mathbb{P}\left(  \left(  \mathbb{E}\left[  (1+T_{\infty
})^{\kappa}-T_{\infty}^{\kappa}\right]  \right)  ^{1/\kappa}\,\eta_{\kappa
}\leq(c_{\kappa}z)^{-1/\kappa}\right) \\
&  =\mathbb{P}\left(  1/\left(  c_{\kappa}\mathbb{E}[(1+T_{\infty})^{\kappa
}-T_{\infty}^{\kappa}]\eta_{\kappa}^{\kappa}\right)  \leq z\right)  \,,
\end{align*}
\noindent where $\eta_{\kappa}$ has characteristic function
\eqref{eq:chfct vs2}.\hfill$\square$

\subsection*{Proof of Lemma \ref{lem: score}}

The results for the score hold by using Lemma \ref{lem: N divided by t conv}%
(i) and establishing, $T^{-1/2}S_{T}=T^{-1/2}S_{\left(  T/\mu\right)
}+o_{\mathbb{P}}\left(  1\right)  $, where $S_{\left(  v\right)  }=\sum
_{i=1}^{\left\lfloor v\right\rfloor }\xi_{i}$. To see that $T^{-1/2}%
S_{T}-T^{-1/2}S_{\left(  T/\mu\right)  }=o_{\mathbb{P}}\left(  1\right)  $,
note that{ for every $M,\delta>0$},
\begin{align*}
\mathbb{P}\left(  T^{-1/2}\left[  S_{T}-S_{\left(  T/\mu\right)  }\right]
>M\right)   &  =\mathbb{P}\left(  T^{-1/2}\left[  S_{T}-S_{\left(
T/\mu\right)  }\right]  >M,\text{ }|N_{T}/T-1/\mu|>\delta\right) \\
&  +\mathbb{P}\left(  T^{-1/2}\left[  S_{T}-S_{\left(  T/\mu\right)  }\right]
>M,\text{ }|N_{T}/T-1/\mu|\leq\delta\right) \\
&  =K_{1}+K_{2}%
\end{align*}
Here, $K_{1}\leq\mathbb{P}\left(  |N_{T}/T-\mu|>\delta\right)  \rightarrow0$,
while, by stationarity,
\begin{align*}
K_{2}  &  \leq\mathbb{P}\left(  T^{-1/2}\max_{T(1/\mu+\delta)\leq s\leq
T(1/\mu+\delta)}|S_{\left(  s\right)  }-S_{\left(  T/\mu\right)  }|>M\right)
\\
&  \leq2\,\mathbb{P}\left(  T^{-1/2}\max_{u\leq T\,\delta}|S_{\left(
u\right)  }|>{M}\right)  \rightarrow2\,\mathbb{P}\left(  \max_{s\leq\delta
}|B(s)|>M/2\right)  \,,
\end{align*}
{ as $T\rightarrow\infty$, where $B$ is a Brownian motion. The right-hand side
converges to zero as $\delta\rightarrow0$. The result for the score then holds
as }$T^{-1/2}S_{\left(  T/\mu\right)  }\overset{d}{\rightarrow}\left(
\tau\Omega/\mu\right)  ^{1/2}{\mathbf{Z}}$ by standard application of a CLT
for martingale differences.

Turning to the information, then b{y the ergodic theorem and as $N_{T}%
\overset{\mathrm{a.s.}}{\rightarrow}\infty$ it follows that $N_{T}^{-1}%
I_{T}\overset{\mathrm{a.s.}}{\rightarrow}\Omega$; see Embrechts et al.~(1997),
Lemma~2.5.3. On the other hand, we have by
Lemma~\ref{lem: N divided by t conv}(i), $N_{T}/T\overset{\mathrm{a.s.}%
}{\rightarrow}1/\mu$. Thus $T^{-1}I_{T}\overset{\mathrm{a.s.}}{\rightarrow
}\Omega/\mu$.}\hfill$\square$

\subsection*{Proof of Lemma \ref{lem: score for k < 1}}

Write
\[
I_{T}(\theta)=-\dfrac{\partial^{2}L_{T}(\theta)}{\partial\theta\partial
\theta^{\prime}}=\sum_{i=1}^{N_{T}}\left(  2\,\varepsilon_{i}\,\dfrac{\psi
_{i}(\theta_{0})}{\psi_{i}(\theta)}-1\right)  {\mathbf{v}}_{i}(\theta
){\mathbf{v}}_{i}(\theta)^{\prime},
\]
where ${\mathbf{v}}_{i}\left(  \theta\right)  =(1,x_{i-1})^{\prime}/\psi
_{i}(\theta)$. The summands constitute a strictly stationary ergodic sequence
with values in the space $\mathbb{C}$ of continuous functions of $\theta$ in a
neighbourhood $\mathcal{N}\left(  \theta_{0}\right)  $ of $\theta_{0}$
equipped with the uniform distance. It is not difficult to see that the
sup-norm of these summands has finite expected value. Therefore the summands
obey the ergodic theorem in $\mathbb{C}$; see Theorem 2.1 in Section~4.2 of
Krengel~(1985). Since $N_{T}\overset{\mathrm{a.s.}}{\rightarrow}\infty$ we
conclude that uniformly over $\theta\in\mathcal{N}(\theta_{0})$,
\[
N_{T}^{-1}I_{T}(\theta)\overset{\mathrm{a.s.}}{\rightarrow}\mathbb{E}%
\Big[\Big(2\,\dfrac{\psi_{1}(\theta_{0})}{\psi_{1}(\theta)}-1\Big){\mathbf{v}%
}_{1}(\theta){\mathbf{v}}_{1}(\theta)^{\prime}\Big]\,\,.
\]
Moreover, together with Lemma~\ref{lem: N divided by t conv} (iv) we conclude
that, uniformly over $\theta\in\mathcal{N}(\theta_{0})$,
\[
T^{-\kappa}I_{T}(\theta)=(N_{T}/T^{\kappa})(N_{T}^{-1}I_{T}(\theta
))\overset{d}{\rightarrow}W\left(  \theta\right)  =\lambda_{\kappa
}\,\mathbb{E}\Big[\Big(2\,\dfrac{\psi_{1}(\theta_{0})}{\psi_{1}(\theta
)}-1\Big){\mathbf{v}}_{1}(\theta){\mathbf{v}}_{1}(\theta)^{\prime}\Big]\,\,.
\]
{ If $\theta-\theta_{0}=O(T^{-\kappa/2})$ then we also have
\[
\sup_{\theta}\left\vert T^{-\kappa}\left(  I_{T}\left(  \theta\right)
-I_{T}(\theta_{0})\right)  \right\vert =\left(  N_{T}/T^{\kappa}\right)
\sup_{\theta}\left\vert N_{T}^{-1}(I_{T}(\theta)-I_{T}(\theta_{0}))\right\vert
\rightarrow0\text{ ,}%
\]
in probability. Thus we verified conditions C1 and C2 in Sweeting (1980) and,
in turn, Theorem 1 applies, yielding}
\[
\left(  T^{-\kappa/2}S_{T}(\theta)\,,T^{-\kappa}I_{T}(\theta)\right)
\overset{d}{\rightarrow}\left(  \left(  \lambda_{\kappa}\Omega\right)
^{1/2}{\mathbf{Z}}\,,\lambda_{\kappa}\Omega\right)
\]
{for a} bivariate{ }standard Gaussian vector $\mathbf{Z}$ independent of
$\lambda_{\kappa}$.\hfill$\square$

\subsection*{Proof of Theorem \ref{thm: new QMLE}}

The result follows by applications of Lemmas 11 and 12 in Kristensen and
Rahbek (2010).{ With the notation there, }set $Q_{T}\left(  \theta\right)
=T^{-1}L_{T}\left(  \theta\right)  $, $U_{T}=1$, and $v_{T}=T$; then
conditions (i),(ii) and (iv) of Lemmas 11 and 12 in Kristensen and Rahbek
(2010) hold by Lemma \ref{lem: score} above, as $\partial^{2}Q_{T}\left(
\theta_{0}\right)  /\partial\theta\partial\theta^{\prime}\overset
{p}{\rightarrow}\Omega/\mu$, and $\partial Q_{T}\left(  \theta_{0}\right)
/\partial\theta\overset{d}{\rightarrow}\left(  \tau\Omega/\mu\right)
^{1/2}{\mathbf{Z}}$. Next, consider condition (iii) of Lemma 11 in Kristensen
and Rahbek (2010); see also Jensen and Rahbek (2004). It follows that in a
compact neighborhood $\mathcal{U}\left(  \theta_{0}\right)  $ of $\theta_{0}$,%
\begin{align}
\sup_{\theta\in\mathcal{U}\left(  \theta_{0}\right)  }\left\vert \partial
^{3}Q_{T}\left(  \theta\right)  /\partial\alpha^{3}\right\vert  &  \leq
T^{-1}\sum_{i=1}^{N_{T}}\left[  2\frac{x_{i}x_{i-1}^{3}}{\psi_{i}^{4}\left(
\theta\right)  }+3\left(  2\frac{x_{i}x_{i-1}^{3}}{\psi_{i}^{4}\left(
\theta\right)  }+\frac{x_{i-1}^{3}}{\psi_{i}^{3}\left(  \theta\right)
}\right)  \right] \label{third}\\
&  \leq\operatorname*{const}\text{ }T^{-1}\sum_{i=1}^{N_{T}}\left(
1+\varepsilon_{i}\right)  \text{,}\nonumber
\end{align}
and condition (iii) holds as $N_{T}/T$ and $T^{-1}\sum_{i=1}^{N_{T}%
}\varepsilon_{i}$ are $O_{\mathbb{P}}\left(  1\right)  $. For the remaining
third-order derivatives similar arguments apply.\hfill$\square$

\subsection*{Proof of Theorem \ref{Theorem EACD MLE kappa <1}}

Similar to the proof of Theorem \ref{thm: new QMLE}, set $Q_{T}\left(
\theta\right)  =T^{-\kappa}L_{T}\left(  \theta\right)  $, $U_{T}=1$, and
$v_{T}=T^{\kappa}$. As there, conditions (i),(ii) and (iv) of Lemmas 11 and 12
in Kristensen and Rahbek (2010)\ hold by Lemma \ref{lem: score for k < 1}.
Likewise as in (\ref{third}),
\begin{equation}
\sup_{\theta\in\mathcal{U}\left(  \theta_{0}\right)  }\left\vert \partial
^{3}Q_{T}\left(  \theta\right)  /\partial\alpha^{3}\right\vert \leq
\operatorname*{const}\text{ }T^{-\kappa}\sum_{i=1}^{N_{T}}\left(
1+\varepsilon_{i}\right)  \text{.}\nonumber
\end{equation}
and condition (iii) holds since $N_{T}/T^{\kappa}$ and $T^{-\kappa}\sum
_{i=1}^{N_{T}}\varepsilon_{i}$ are $O_{\mathbb{P}}\left(  1\right)  $%
.\hfill$\square$

\subsection*{Proof of Corollary \ref{Cor t}}

For $\kappa>1$ the result follows from Theorem \ref{thm: new QMLE} by standard
arguments. For $\kappa<1$, the result holds by the proof of Theorem
\ref{Theorem EACD MLE kappa <1} and Corollary 1 in Sweeting (1980).\hfill
$\square$

\section{The remainder term}

\label{Appendix B}

We noticed in Section \ref{sec asymptotics} that the likelihood function in
\eqref{eq: EACD likelihood} in addition to the observed durations
$\{x_{i}\}_{i=1}^{N_{T}}$ in $\left[  0,T\right]  $ misses out on a term for
$t_{N_{T}}<T$, i.e., the term containing the information about no events
occurring in $(t_{N_{T}},T]$ is ignored. Thus, strictly speaking, $\hat
{\theta}_{T}$ is not the MLE. While it is common practice to ignore this
likelihood contribution in the ACD\ literature (as standard ARCH software is
typically applied for estimation), this term is usually included in the
related point process literature; cf. Daley and Vere-Jones (2008). By using
the point process representation of the ACD process, as originally noted in
Engle and Russell (1998), it holds that the ACD\ conditional intensity
$\lambda(t|\left\{  x_{i}\right\}  _{i=0}^{N_{t}})=1/\psi_{N_{t}+1}$ which
implies that the remainder term missing in (\ref{eq: EACD likelihood}),
$R_{T}\left(  \theta\right)  $ say, is given by%
\begin{equation}
R_{T}\left(  \theta\right)  =-\tfrac{T-t_{N_{T}}}{\psi_{N_{T}+1}}\text{.}
\label{eq: remainder}%
\end{equation}
We now establish that under geometric ergodicity $R_{T}\left(  \theta\right)
$ is asymptotically negligible.

\begin{lemma}
\label{lem: remainder}Consider the remainder term $R_{T}\left(  \theta\right)
$ given by \eqref{eq: remainder}. Then for $\left\{  x_{i}\right\}  $ strictly
stationary and geometrically ergodic, we have $R_{T}\left(  \theta\right)
T^{-1/2}\overset{p}{\rightarrow}0$ for $\kappa>1$, and $R_{T}\left(
\theta\right)  T^{-\kappa/2}\overset{p}{\rightarrow}0$ for $\kappa<1$.
\end{lemma}

\noindent\textsc{Proof.}\ With $\nu_{T}=N_{T}+1$ we have $t_{N_{T}}%
<T<t_{\nu_{T}}$,
\[
\max\left(  t_{\nu_{T}}-T,T-t_{N_{T}}\right)  <x_{\nu_{T}}\text{,}%
\]
and $R_{T}(\theta)=(T-t_{N_{T}})/{\psi_{\nu_{T}}}\leq\varepsilon_{\nu_{T}}$.

For $\kappa>1$, $\lim_{T\rightarrow\infty}\nu_{T}/T=\lim_{T\rightarrow\infty
}N_{T}/T=1/\mu$ a.s. and for $\gamma>0$
\begin{align*}
\mathbb{P}(\varepsilon_{\nu_{T}}>\sqrt{T})  &  =\mathbb{P}(\varepsilon
_{\nu_{T}}>\sqrt{T}\,,\left\vert \nu_{T}/T-1/\mu\right\vert >\gamma
)+\mathbb{P}(\varepsilon_{\nu_{T}}>\sqrt{T},\left\vert \nu_{T}/T-1/\mu
\right\vert \leq\gamma)\\
&  =I_{1}+I_{2}\text{.}%
\end{align*}
For every $\gamma>0$, $I_{1}\leq\mathbb{P}\left(  \left\vert \nu_{T}%
/T-1/\mu\right\vert >\gamma\right)  \rightarrow0$, while for small $\gamma$,
\begin{align*}
I_{2}  &  \leq\mathbb{P}\left(  \max_{(1/\mu-\gamma)\leq s\leq T(1/\mu
+\gamma)}\varepsilon_{s}>\sqrt{T}\right)  \leq\mathbb{P}\left(  \max
_{s\leq3\gamma T}\varepsilon_{s}>\sqrt{T}\right) \\
&  \leq3\gamma T\,\mathbb{P}(\varepsilon>\sqrt{T})\rightarrow0\,.
\end{align*}

Next, for $\kappa<1$, $v_{T}/T^{\kappa}$ and $N_{T}/T^{\kappa}$ converge in
distribution to $\lambda_{\kappa}$, and hence%
\begin{align*}
\mathbb{P}(\varepsilon_{\nu_{T}}>T^{\kappa/2})  &  =\mathbb{P}(\varepsilon
_{\nu_{T}}>T^{\kappa/2},\nu_{T}/T^{\kappa}>M)+\mathbb{P}(\varepsilon_{\nu_{T}%
}>T^{\kappa/2},\nu_{T}/T^{\kappa}\leq M)\\
&  =K_{1}+K_{2}\text{.}%
\end{align*}
Here $K_{1}\leq\mathbb{P}\left(  \nu_{T}/T^{\kappa}>M\right)  $ is arbitrarily
small for large $M$ as $T\rightarrow\infty$ while
\[
K_{2}\leq\mathbb{P}\Big(\max_{s\in3MT^{\kappa}}\varepsilon_{s}>T^{\kappa
/2}\Big)\leq3\,M\,T^{\kappa}\,\mathbb{P}(\varepsilon>T^{\kappa/2}%
)\leq\mathrm{const}\,T^{\kappa}\,\mathbb{P}(\varepsilon>T^{\kappa
/2})\rightarrow0\,,
\]
as desired. $\hfill\square$

\end{document}